\documentclass[letterpaper, 10 pt, journal, twoside]{IEEEtran}

\IEEEoverridecommandlockouts                              

\usepackage[dvips]{graphics} 
\usepackage{epsfig} 
\usepackage{mathptmx} 
\usepackage{times} 
\usepackage{amsmath} 
\usepackage{amssymb}  

\usepackage[subrefformat=parens]{subcaption}
\usepackage{amsfonts}
\usepackage{mathrsfs}

\title{Sensitivity Analysis of Cascaded Quantum Feedback Amplifier}

\author{Yu Yokotera and Naoki Yamamoto
\thanks{Y. Yokotera and N. Yamamoto are with the Department of Applied Physics and Physico-Informatics, Keio University, Yokohama 223-8522, Japan (e-mail: y-yokotera@z6.keio.jp; yamamoto@appi.keio.ac.jp). }
\thanks{This work was supported by JST PRESTO under Grant JPMJPR166A.}
}

\begin{document}

\pagestyle{empty} 
\maketitle
\thispagestyle{empty} 


\begin{abstract}

Quantum amplifier is an essential device in quantum information processing. 
As in the classical (non-quantum) case, its characteristic uncertainty needs to be suppressed by feedback, and in fact such a control theory for a single quantum amplifier has recently been developed. This letter extends this result to the case of cascaded quantum amplifier. In particular, we consider two types of structures: 
the case where controlled amplifiers are connected in series, and the case where a single feedback control is applied to the cascaded amplifier. 
Then, we prove that the latter is better in the sense of sensitivity to the uncertainty. 
A detailed numerical simulation is given to show actual performance of these two feedback schemes. 

\end{abstract}



\section{Introduction}
\label{sec:1}

\IEEEPARstart{A}{mplifiers} are essential in modern technology \cite{Graeme1971}. 
Note that this device is not used in a stand-alone fashion, because its amplification gain cannot 
be exactly specified due to unavoidable characteristic uncertainty. 
Actually the amplified output signal produced from a bare amplifier can be largely distorted, and 
eventually the performance of signal processing is degraded. 
Black discovered that feedback control resolves this issue \cite{Black1934}, which has been 
further investigated in depth \cite{Blackman1943, Bode1945}. 
This feedback amplification method, which is now known as one of the most successful examples 
of control theory, has made a significant contribution to the development of the today's electronic 
technologies.

\begin{figure}[!t]
\begin{center}
\includegraphics[width=3.2cm]{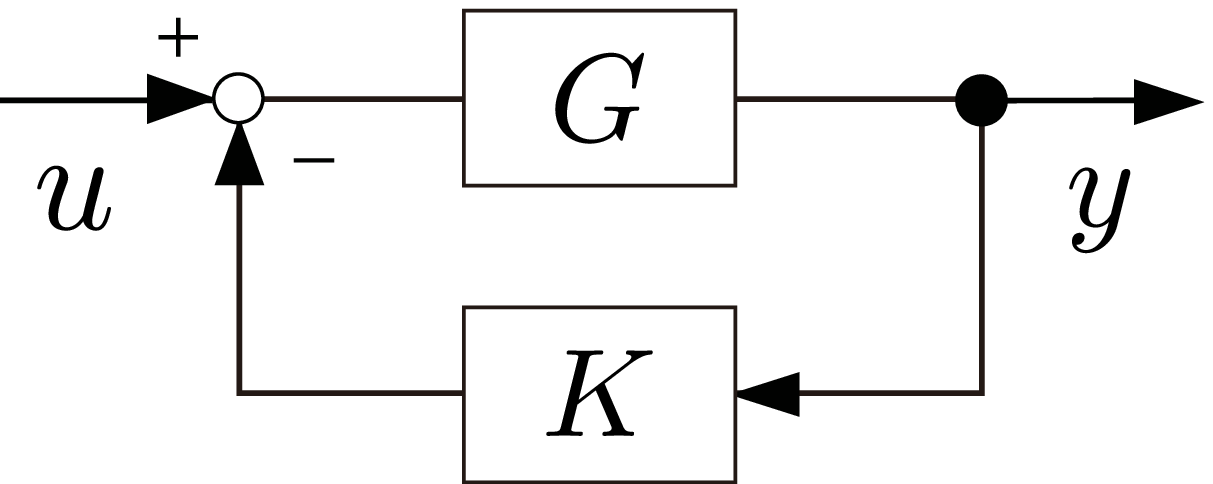}
\caption{Feedback control for a classical amplifier.\newline}
\label{Fig1_classicalamp}
\includegraphics[width=0.9\hsize]{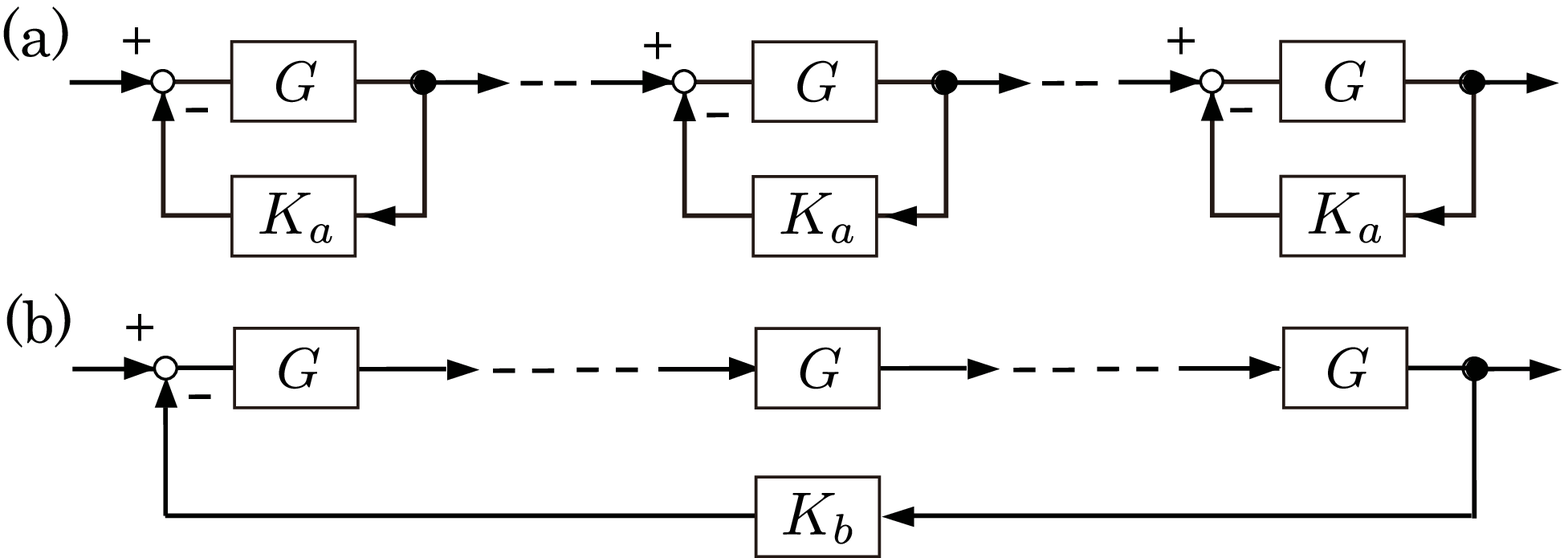}
\caption{Two basic feedback configurations for cascaded classical amplifier; 
type-a and type-b.}
\label{Fig2_cascadedcamp}
\end{center}
\end{figure}

The idea of classical (non-quantum) feedback amplification is as follows. 
Figure~\ref{Fig1_classicalamp} shows a system composed of a single amplifier with gain $G$ 
and another system (called the controller) with gain $K$. 
A simple calculation yields 
\begin{align}
      y=G^{\rm fb}u, ~~G^{\rm fb}=\frac{G}{1+GK}.
\label{eq_classicalFBamp}
\end{align}
Therefore, in the limit $|G| \to \infty$, the closed-loop gain becomes $G^{\rm fb} \to 1/K$. 
This means that the robust amplification is realized by taking a passive and attenuating 
controller, such as a resistor, because the characteristic change in $K$ of those passive devices 
is in general quite small.

A single amplifier does not always provide sufficient gain and bandwidth due to the 
gain-bandwidth constraint, and thus cascaded amplifiers are often used in practice to satisfy 
the required performance \cite{Graeme1971}. 
Surely feedback stabilization is needed in this case as well, but it is not obvious how to 
construct a feedback configuration for such a multi-component network. 
In the classical control theory, as the most basic study, two types of feedback configurations 
depicted in Fig.~\ref{Fig2_cascadedcamp} were first investigated. 
The type-a scheme shown in Fig.~\ref{Fig2_cascadedcamp}(a) is the cascade connection 
of the feedback-controlled amplifiers, and in the type-b scheme shown in (b), a single feedback 
loop is constructed for the cascaded amplifier. 
In \cite{Horrocks1990}, it was shown that the type-b scheme is more effective for improving 
the robustness than the type-a.

Turning our attention to the quantum regime, the quantum amplifier 
\cite{Louisell1961}--\cite{Caves2012} 
is expected to serve as a fundamental 
device in quantum information science, such as quantum sensing 
\cite{Caves1987}--\cite{Abdo2014}
and quantum communication 
\cite{Pirandola2017}--\cite{Elemy2017}. 
In practice, the quantum amplifier must be stabilized via feedback as in the classical case. 
In fact one of the authors has developed the theory of feedback stabilization for a {\it single} 
quantum amplifier \cite{Yamamoto2016}. 
It is thus important to extend the theory to the case of cascaded quantum amplifier 
\cite{Yan2012}--\cite{Wang2013}, 
which has not been yet established; 
in particular, analyzing proper quantum versions of the above-described two classical feedback 
configurations should be an important basic study along this research direction. 
The contribution of this letter is to prove that a quantum version of the classical type-b 
scheme is better than a correspondence to the type-a, in the sense of the robustness. 
Note that, although this is the same conclusion as the classical one, the proof is non-trivial, 
because the quantum amplifier is essentially a multi-input and multi-output (MIMO) device 
and eventually the analysis becomes much more involved than the classical case, as will 
be shown in the letter.

This letter is organized as follows. 
Section \ref{sec:2} is devoted to some preliminaries. 
In Section \ref{sec:3}, we prove the main result. 
Section \ref{sec:4} gives a detailed numerical simulation to show the robustness and stability 
of the controlled amplifiers.


\section{Preliminaries}\label{sec:2}

\subsection{Sensitivity Function and Cascaded Classical Amplifier}
\label{sec:2.1}

Here we aim to quantify the robustness of the controlled amplifier described in Section~\ref{sec:1}. 
Suppose that a small characteristic change $\Delta G$ occurs in the gain as $G \to G+\Delta G$. 
Then the closed-loop gain \eqref{eq_classicalFBamp} changes to 
$G^{\rm fb}+\Delta G^{\rm fb}$. 
The {\it sensitivity function} of $G^{\rm fb}$ with respect to $G$ is defined as 
\begin{align}
      S=\frac{\,\, \Delta G^{\rm fb}/G^{\rm fb}}{\Delta G/G}.
\label{eq_sensitivity def}
\end{align} 
Now the small deviation $\Delta G^{\rm fb}$ is calculated as 
\[
\Delta G^{\rm fb}
    =\frac{G+\Delta G}{1+(G+\Delta G)K}-\frac{G}{1+GK} \approx \frac{\Delta G}{(1+GK)^2},
\]
thus $S=1/(1+GK)$. 
Therefore, the open-loop gain $GK$ should be carefully designed so that $|S| < 1$ while 
retaining the stability of the closed-loop system.

Next we consider the cascaded feedback amplifiers shown in 
Fig.~\ref{Fig2_cascadedcamp} \cite{Horrocks1990}, which in both cases are 
composed of $N$ identical classical amplifiers. 
In the type-a scheme, the same feedback controller with gain $K_{a}$ is applied to 
each amplifier, and in the type-b scheme, the output of the terminal amplifier is fed back to 
the first one through the single controller with gain $K_{b}$. 
The overall gains are given by 
\begin{equation*}
    G_{a}^{\rm fb} = (G^{\rm fb})^{N}=\frac{G^N}{(1+GK_{a})^N}, ~~
    G_b^{\rm fb} = \frac{G^N}{1+G^{N}K_b}.
\end{equation*} 
Now suppose that the small change $G \to G+\Delta G$ occurs in one of the amplifiers, 
say, the $j$-th amplifier. 
Then the fluctuations of $G_{a}^{\rm fb}$ and $G_{b}^{\rm fb}$ are calculated as follows;
\begin{align*}
\Delta G_{a}^{\rm fb}
&=\frac{(G+\Delta G)G^{N-1}}{[1+(G+\Delta G)K_{a}](1+GK_{a})^{N-1}}-\frac{G^N}{(1+GK_{a})^N} \\
&\approx \frac{G^{N-1}\Delta G}{(1+GK_{a})^{N+1}},\\
\Delta G_{b}^{\rm fb}
&=\frac{(G+\Delta G)G^{N-1}}{1+(G+\Delta G)G^{N-1}K_{b}}-\frac{G^N}{1+G^{N}K_{b}}
\approx \frac{G^{N-1}\Delta G}{(1+G^{N}K_{b})^{2}}.
\end{align*} 
From Eq.~\eqref{eq_sensitivity def}, the sensitivity functions are given by
\begin{equation}
\label{C sensitivity fn}
     S_{a} = 1/(1+GK_{a}),~~
     S_{b} = 1/(1+G^{N}K_{b}).
\end{equation}
Then, if the gains of both of the controlled systems are equal and these are smaller than 
the gain of the non-controlled cascaded amplifier, 
i.e., $|G^{\rm fb}_a|=|G^{\rm fb}_b| <|G|^N$, we have 
\begin{equation*}
    \frac{|S_{b}|}{|S_{a}|}=\frac{1}{|1+GK_{a}|^{N-1}}<1.
\end{equation*}
Thus the type-b feedback scheme has a better performance than the type-a scheme 
in the sense of sensitivity.


\subsection{Quantum Amplifier and Feedback Stabilization}
\label{sec:2.2}

In this letter we consider the {\it phase-preserving linear quantum amplifier} 
\cite{Louisell1961}--\cite{Caves2012}, 
which is simply called the ``amplifier" in what follows. 
Let $b(t)$ be a field annihilation operator called the {\it signal}; 
$b(t)$ has the meaning of a complex amplitude of the field and satisfies the canonical 
commutation relation (CCR) $b(t)b^\dagger(t') - b^\dagger(t') b(t)=\delta(t-t')$, where 
$b^{\dag}(t)$ represents the Hermitian conjugate of $b(t)$. 
The amplifier transforms $b(t)$ to $\tilde{b}(t)=gb(t) +\sqrt{g^2 -1}\,d^{\dag}(t)$, where $d(t)$ 
is an additional field annihilation operator called the {\it idler}, which is necessary to preserve 
the CCR of $\tilde{b}(t)$. 
Also $g>1$ is the amplification gain.

\begin{figure}[!t]
\begin{center}
\includegraphics[width=8.8cm]{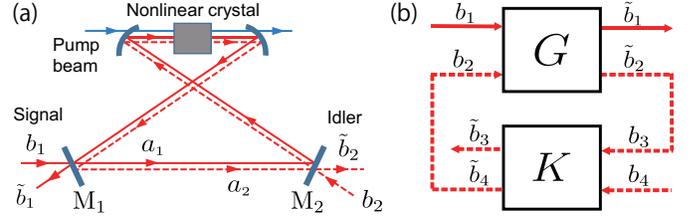}
\caption{(a) Non-degenerate parametric amplifier. 
(b) Feedback configuration for a single quantum amplifier.}
\label{Fig3_NDPA_FBamp}
\end{center}
\end{figure}

In quantum optics the non-degenerate parametric amplifier (NDPA) \cite{Ou1992} shown in 
Fig.~\ref{Fig3_NDPA_FBamp}(a) is often used. 
This is an optical cavity with two inputs $b_{1}$ (signal) and $b_{2}$ (idler). 
The corresponding internal cavity modes $a_{1}$ and $a_{2}$ couple with each other at the 
pumped nonlinear crystal. 
In the rotating-frame at the half of input laser frequency, the dynamical equations of the NDPA 
under ideal setup (i.e., zero-detuned and no-loss) are given by
\begin{align*}
      &\dot{a}_{1}=-\frac{\kappa}{2} a_{1} 
              +\varepsilon a_{2}^{\dag}-\sqrt{\kappa}b_{1}, ~
        \dot{a}_{2}^{\dag}=-\frac{\kappa}{2} a_{2}^{\dag}
              +\varepsilon a_{1}-\sqrt{\kappa}b_{2}^{\dag}, \\
&\tilde{b}_{1}=\sqrt{\kappa}a_{1}+b_{1}, ~~
   \tilde{b}_{2}^{\dag}=\sqrt{\kappa}a_{2}^{\dag}+b_{2}^{\dag},
\end{align*}
where $\kappa$ is the cavity damping rate and $\varepsilon$ is the strength of nonlinearity. 
(The mirror ${\rm M}_i$ is partially transmissive for $a_{i}$ but perfectly reflective for the 
other cavity mode.)

In the Laplace domain, the amplified output signal $\tilde{b}_{1}$ is, together with the amplified 
idler $\tilde{b}_{2}$, represented as 
\begin{equation}
\label{NDPA gain}
    \left[\begin{array}{c}
      \tilde{b}_{1}(s)\\
      \tilde{b}_{2}^{\dag}(s)\\
     \end{array}\right] 
       =\left[\begin{array}{cc}
            g_{1}(s) & g_{2}(s) \\
            g_{2}(s) & g_{1}(s) \\
         \end{array}\right]
         \left[\begin{array}{c}
            b_{1}(s)\\
            b_{2}^{\dag}(s)\\
         \end{array}\right],
\end{equation}
where $g_{1}(s)=(s^{2}-\kappa^{2}/4-\varepsilon^2) / D(s)$ and $g_{2}(s)=-\kappa \varepsilon/D(s)$ 
are the transfer functions with $D(s)=s^{2}+\kappa s+ \kappa^{2}/4-\varepsilon^2$. 
Note that $|g_{1}(i\omega)|^2-|g_{2}(i\omega)|^2=1,~\forall \omega$ holds to satisfy the 
CCR of the output, represented by 
$\tilde{b}(i\omega)\tilde{b}^\dagger(i\omega') - \tilde{b}^\dagger(i\omega')\tilde{b}(i\omega)
=\delta(\omega-\omega')$ in the Fourier domain. 
Also the characteristic equation $D(s)=0$ yields the stability condition 
$0< x=2\varepsilon/\kappa < 1$. 
The gain at the center frequency satisfies $|g_{1}(0)|=(1+x^2)/|1-x^2| \to \infty$ as $x \to 1-0$.

Here we review the general feedback method for a single quantum amplifier \cite{Yamamoto2016}. 
The general linear quantum amplifier is represented in the Laplace domain as 
\cite{Gough2010}: 
\begin{align*}
     \left[\begin{array}{c}
          \tilde{b}_{1}(s)\\
          \tilde{b}_{2}^{\dag}(s)\\
     \end{array}\right ] 
   =G(s)
      \left[\begin{array}{c}
          b_{1}(s)\\
          b_{2}^{\dag} (s)\\
      \end{array}\right], ~ 
   G(s)=
      \left[\begin{array}{cc}
         G_{11}(s) & G_{12}(s) \\
         G_{21}(s) & G_{22}(s) \\
      \end{array}\right], 
\end{align*} 
where $|G_{11}(i\omega)|^2 - |G_{12}(i\omega)|^2 = |G_{22}(i\omega)|^2 - |G_{21}(i\omega)|^2 = 1$ 
and $G_{21}(i\omega)G_{11}^*(i\omega) - G_{22}(i\omega)G_{12}^*(i\omega)=0$ hold 
for all $\omega$. 
As for the controller, we take a passive and attenuating quantum system with the following 
input-output relation: 
\begin{align*}
\hspace{-0.08cm}
   \left[\begin{array}{c}
     \tilde{b}_{3}^{\dag}(s)\\
     \tilde{b}_{4}^{\dag}(s)\\
   \end{array}\right ] 
   =K(s)
     \left[\begin{array}{c}
        b_{3}^{\dag} (s)\\
        b_{4}^{\dag} (s)\\
     \end{array}\right], ~   
   K(s)=\left[\begin{array}{cc}
       K_{11}(s) & K_{12}(s) \\
       K_{21}(s) & K_{22}(s) \\
     \end{array}\right]. 
\end{align*} 
$K^{\dag}(i\omega)K(i\omega)=I,~\forall \omega$ holds to satisfy the CCR in both 
$\tilde{b}_{3}$ and $\tilde{b}_{4}$. 
These two systems are connected through the feedback $b_{2}=\tilde{b}_{4}$ and 
$b_{3}=\tilde{b}_{2}$, as shown in Fig.~\ref{Fig3_NDPA_FBamp}(b). 
The input-output relation of the closed-loop system is given by
\begin{equation}
\label{closed-loop G general}
    \left[\begin{array}{c}
      \tilde{b}_{1}(s)\\
      \tilde{b}_{3}^{\dag}(s)\\
     \end{array}\right] 
       =\left[\begin{array}{cc}
            G^{\rm fb}_{11}(s) & G^{\rm fb}_{12}(s) \\
            G^{\rm fb}_{21}(s) & G^{\rm fb}_{22}(s) \\
         \end{array}\right]
         \left[\begin{array}{c}
            b_{1}(s)\\
            b_{4}^{\dag}(s)\\
         \end{array}\right],
\end{equation}
where 
\begin{align*}
    G^{\rm fb}_{11}&=[G_{11}-K_{21}(G_{11}G_{22}-G_{12}G_{21})] / (1-G_{22}K_{21}),  \\ 
    G^{\rm fb}_{12}&=G_{12}K_{22} / (1-G_{22}K_{21}), ~~
    G^{\rm fb}_{21}=G_{21}K_{11} / (1-G_{22}K_{21}), \\
    G^{\rm fb}_{22}&=[ K_{12}+G_{22}(K_{11}K_{22}-K_{12}K_{21})] / (1-G_{22}K_{21}).  
\end{align*} 
Then $|G^{\rm fb}_{11}(s)| \to 1/|K_{21}(s)| > 1$ holds in the high-gain limit $|G_{11}| \to \infty$, 
meaning that the amplification process can be made robust by feedback as in the classical case.


\section{Cascaded quantum feedback amplifier}
\label{sec:3}

\begin{figure}[!t]
\begin{center}
\includegraphics[width=8.5cm]{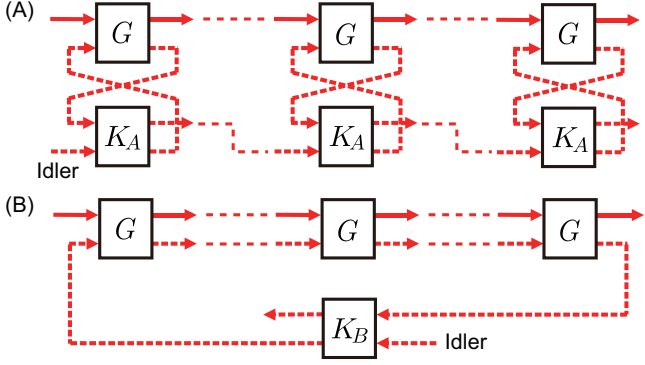}
\caption{
Two basic feedback configurations for cascaded quantum amplifier; 
type-A and type-B. 
}
\label{fig_cascadeQamp}
\end{center}
\end{figure}

In this section we show the quantum version of the classical cascade amplification theory 
given in Section~\ref{sec:2.1}. 
First note that, because the quantum amplifier is an MIMO system and hence it essentially 
differs from the classical one, specifying the feedback network composed of amplifiers and 
controllers, which corresponds to the classical one shown in Fig.~\ref{Fig2_cascadedcamp}, 
is a non-trivial problem. 
Here we particularly consider the case where the idler mode of the amplifier can be used, 
in addition to the signal mode, to construct the feedback network; 
actually in the standard experiments of quantum optics \cite{Ou1992} and superconductivity 
\cite{Yurke1988}, the idler mode is artificially implemented and is thus accessible. 
In this formulation, reasonable quantum versions of the classical feedback networks are 
illustrated in Fig.~\ref{fig_cascadeQamp}; 
the type-A and type-B schemes correspond to the classical type-a and type-b schemes, 
respectively.
In both cases, the signal-out and the idler-out are connected to the signal-in and the idler-in, 
respectively, and eventually the whole system has only one idler input from outside. 
Note that, if the idler modes are not accessible and only the signal modes can be connected, 
then in both configurations the whole closed-loop system has multiple idler inputs 
and eventually it is subjected to a large noise coming from those idler input channels.

Now the problem is to compare the sensitivity of the two schemes shown in 
Fig.~\ref{fig_cascadeQamp}. 
We tackle this problem under the following setting. 
First, we focus on the gain at the center frequency $\omega=0$. 
Then we consider the quantum amplifier whose transfer function matrix at $\omega=0$ is 
of the form 
\begin{align}
   G(0)=\left[\begin{array}{cc}
               G_{1} & G_{2} \\  
               G_{2} & G_{1} \\
             \end{array}\right], ~~ G_{1}^{2}-G_{2}^{2}=1, ~~G_i\in{\mathbb R}.  
\label{eq_ideal Qamp}
\end{align}
Note that the ideal NDPA with transfer functions \eqref{NDPA gain} indeed fulfills this condition. 
Moreover we suppose that both feedback networks are composed of $N$ identical quantum 
amplifiers characterized by Eq.~\eqref{eq_ideal Qamp}, and that the gain of only the $j$-th 
amplifier changes as $G_{1} \to G_{1} +\Delta G_{1}$ and $G_{2} \to G_{2} +\Delta G_{2}$. 
Lastly, without loss of generality, the transfer function matrix of the controller at $\omega=0$ can be set to: 
\begin{align*}
    K_\bullet(0) = \left[\begin{array}{cc}
                 K_{1\bullet} & K_{2\bullet} \\  
                 -K_{2\bullet} & K_{1\bullet} \\
               \end{array} \right ] , ~ K_{1\bullet}^2+K_{2\bullet}^2=1, ~~K_{i \bullet}\in{\mathbb R}, 
\end{align*}
where $\bullet=A, B$; 
i.e., $K_A(0)$ and $K_B(0)$ are applied to the type-A and the type-B schemes, respectively.

First, we derive the overall gain for the type-A scheme. 
From Eq.~\eqref{closed-loop G general}, each feedback-controlled amplifier has the 
following transfer function matrix: 
\begin{align*}
G^{\rm fb}(0)
&=\left[\begin{array}{cc}
           G^{\rm fb}_{1} & G^{\rm fb}_{2} \\
           G^{\rm fb}_{2} & G^{\rm fb}_{1}  \\
  \end{array}\right]
  \nonumber \\
&=\frac{1}{1+G_{1}K_{2A}}
\left[\begin{array}{cc}
           G_{1}+K_{2A} & G_{2}K_{1A}  \\
           G_{2}K_{1A} & G_{1}+K_{2A}  \\
      \end{array} \right]. 
\end{align*}
This matrix can be diagonalized using the orthogonal matrix $P=1/\sqrt{2}[1, 1;1, -1]$ as follows;
\begin{align*}
     P^{-1}G^{\rm fb}(0)P
        = {\rm diag}\{\lambda^{\rm fb}_{+}, ~
               \lambda^{\rm fb}_{-} \}
        =\left[\begin{array}{cc}
              \lambda^{\rm fb}_{+} & 0 \\  [3pt] 
              0 & \lambda^{\rm fb}_{-}  \\
           \end{array}\right ],
\end{align*}
where $\lambda^{\rm fb}_{\pm}=(G_1+K_{2A} \pm G_{2}K_{1A})/(1+G_1K_{2A})$. 
The overall transfer function matrix is the $N$ product of $G^{\rm fb}(0)$;
\begin{align*}
    G_{A}^{\rm fb} & \equiv 
          \left[\begin{array}{cc}
    G^{\rm fb}_{1A} & G^{\rm fb}_{2A} \\[3pt] 
    G^{\rm fb}_{2A} & G^{\rm fb}_{1A} \\
          \end{array}\right ]
       = [G^{\rm fb}(0)]^N \\
     &=\frac{1}{2}
            \left[\begin{array}{cc}
              (\lambda^{\rm fb}_{+})^N + (\lambda^{\rm fb}_{-})^N 
          & (\lambda^{\rm fb}_{+})^N - (\lambda^{\rm fb}_{-})^N \\[3pt]
             (\lambda^{\rm fb}_{+})^N -  (\lambda^{\rm fb}_{-})^N 
          & (\lambda^{\rm fb}_{+})^N + (\lambda^{\rm fb}_{-})^N \\
            \end{array}\right ].
\end{align*}
The gain of interest is the (1,1) element of $G_{A}^{\rm fb}$, i.e., $G^{\rm fb}_{{\scriptstyle 1A}}$. 
Now the characteristic changes $G_{1} \to G_{1} +\Delta G_{1}$ and 
$G_{2} \to G_{2} +\Delta G_{2}$ occur; then, using $G_2\Delta G_{2}=G_{1}\Delta G_{1}$, we 
find that the fluctuation of $G^{\rm fb}_{1A}$ is calculated as 
\begin{align*}
\Delta G^{\rm fb}_{1A}
&=\frac{1}{2}\left[
\frac{G_{1}+\Delta G_{1}+K_{2A}+(G_{2}+\Delta G_{2})K_{1A}}{1+(G_{1}+\Delta G_{1})K_{2A}} \right. \\
&~~~~~~~~~~\left. 
-
~\frac{G_{1}+K_{2A}+G_{2}K_{1A}}{1+G_{1}K_{2A}}
\right]
\left( \lambda^{\rm fb}_{+} \right)^{N-1} \\
&~~~~+\frac{1}{2}\left[
\frac{G_{1}+\Delta G_{1}+K_{2A}-(G_{2}+\Delta G_{2})K_{1A}}{1+(G_{1}+\Delta G_{1})K_{2A}} \right. \\
&~~~~~~~~~~\left. 
-
~\frac{G_{1}+K_{2A}-G_{2}K_{1A}}{1+G_{1}K_{2A}}
\right]
\left( \lambda^{\rm fb}_{-} \right)^{N-1} \\
&\approx
\frac{K_{1A} \Delta G_{1}}{2G_{2}(1+G_{1}K_{2A})}
\left[
\left( \lambda^{\rm fb}_{+} \right)^{N}
-\left( \lambda^{\rm fb}_{-} \right)^{N}
\right] .
\end{align*}
As a result, the sensitivity function is represented as 
\begin{equation}
\label{Q sensitivity fn A}
    S_{A}
        =\frac{\,\, \Delta G^{\rm fb}_{1A}/G^{\rm fb}_{1A}}{\Delta G_{1}/G_{1}}
        =\frac{K_{1A}G_{1}}{G_{2}(1+G_{1}K_{2A})}\frac{G^{\rm fb}_{2A}}{G^{\rm fb}_{1A}}.
\end{equation}

Next we consider the type-B scheme, where the single feedback control is applied to the 
series of quantum amplifiers with transfer function matrix \eqref{eq_ideal Qamp}. 
Noting that $G(0)$ is diagonalized in terms of the orthogonal matrix $P$ as 
$P^{-1}G(0)P 
= {\rm diag}\{ \lambda_{+}, ~ \lambda_{-}  \}$ with 
$\lambda_{\pm}=G_{1} \pm G_{2}$, we have 
\begin{align*}
    [G(0)]^N
       = \left[\begin{array}{cc}
            M_1 & M_2  \\ 
            M_2 & M_1 \\
           \end{array}\right]
        =\frac{1}{2}
           \left[\begin{array}{cc}
                \lambda_{+}^N + \lambda_{-}^N 
            & \lambda_{+}^N - \lambda_{-}^N \\[3pt]
               \lambda_{+}^N -  \lambda_{-}^N 
            & \lambda_{+}^N + \lambda_{-}^N \\
           \end{array}\right]. 
\end{align*}
From Eq.~\eqref{closed-loop G general}, the transfer function matrix of the whole closed-loop 
system is then given by 
\begin{align*}
    G_{B}^{\rm fb} & \equiv 
        \left[\begin{array}{cc}
           G^{\rm fb}_{1B} & G^{\rm fb}_{2B} \\ [3pt]
           G^{\rm fb}_{2B} & G^{\rm fb}_{1B}  \\
        \end{array}\right] 
\\ \nonumber
     &=\frac{1}{1+M_{1}K_{2B}}
            \left[\begin{array}{cc}
               M_{1}+K_{2B} & M_{2}K_{1B} \\[3pt] 
               M_{2}K_{1B} & M_{1}+K_{2B}  \\
            \end{array}\right]. 
&\end{align*}
The characteristic change in $G_1$ and $G_2$ induces a small fluctuation in the overall gain, 
$G^{\rm fb}_{1B}$, as follows: 
\begin{align*}
\Delta G^{\rm fb}_{1B}
&=\frac{M_{1}+\Delta M_{1}+K_{2B}}{1+(M_{1}+\Delta M_{1})K_{2B}}-\frac{M_{1}+K_{2B}}{1+M_{1}K_{2B}} \\
&=\frac{K_{1B}^2 \Delta M_{1}}{(1+M_{1}K_{2B})[1+(M_{1}+\Delta M_{1})K_{2B}]}  \\
&=\frac{K_{1B}^2 M_{2} \Delta G_{1}}
            {(1+M_{1}K_{2B})[G_{2}+(M_{1}G_{2}+M_{2}\Delta G_{1})K_{2B}] } \\
&\approx \frac{K_{1B}G^{\rm fb}_{2B}\Delta G_{1}}{G_{2}(1+M_{1}K_{2B})},
\end{align*}
where the following equality is used:
\begin{align}
     G_2 \Delta M_{1} = M_{2} \Delta G_{1}.
\label{eq_fluctuation relationship}
\end{align}
The proof of this equation is given in Appendix. 
Therefore we arrive at the following sensitivity function: 
\begin{equation}
\label{Q sensitivity fn B}
    S_{B}
       =\frac{\,\, \Delta G^{\rm fb}_{1B}/G^{\rm fb}_{1B}}{\Delta G_{1}/G_{1}}
       =\frac{K_{1B}G_{1}}{G_{2}(1+M_{1}K_{2B})} \frac{G^{\rm fb}_{2B}}{G^{\rm fb}_{1B}}.
\end{equation}

Now we show the main result of this letter; 
if the gains of both of the controlled systems are equal and these are smaller than the gain of 
the non-controlled cascaded amplifier, 
i.e., $|G^{\rm fb}_{1A}|=|G^{\rm fb}_{1B}| < |M_1|$, we prove that 
\begin{equation}
\label{main result}
     |S_{B}| < |S_{A}|.
\end{equation}
The proof is given in Appendix. 
Therefore, the type-B feedback scheme is better than the type-A scheme in terms of the 
sensitivity to the characteristic uncertainty $\Delta G_1$.

{\it Remark 1:} 
Here we remark on a difference between the quantum and classical sensitivity functions. 
Because we aim to construct a high-gain feedback controlled amplifier, let us assume 
$|G^{\rm fb}_{1\bullet}|\gg 1$ ($\bullet=A, B$). 
Then, due to $|G^{\rm fb}_{1\bullet}|^2 - |G^{\rm fb}_{2\bullet}|^2 =1$, 
Eqs. \eqref{Q sensitivity fn A} and \eqref{Q sensitivity fn B} are then approximated as 
\[
     S_{A} \approx \frac{K_{1A}G_{1}}{G_{2}(1+G_{1}K_{2A})},~~~
     S_{B} \approx \frac{K_{1B}G_{1}}{G_{2}(1+M_{1}K_{2B})}.
\]
Now further let us take $|G_1|\gg 1$; then, from $G_1^2-G_2^2 =1$, the quantum sensitivity 
function $S_\bullet$ is identical to the classical one \eqref{C sensitivity fn} except for 
$K_{1\bullet}$. 
However, the idea of cascade amplification is to connect many low-gain amplifiers in series 
to realize $|G^{\rm fb}_{1\bullet}|\gg 1$ (e.g., Case 4 in Section~\ref{sec:4}); 
in this case $G_1/G_2$ takes a large value, and eventually $S_\bullet$ can become bigger 
than the classical one or even 1. 
In the classical case, this type of performance degradation does not occur, which is due to 
the increase of $G_1/G_2$; note that this term stems from the CCR constraint on quantum 
mechanical systems.


\section{Stability and sensitivity analysis}
\label{sec:4}

\begin{table}[!t]
\begin{center}
\caption{Nominal parameters and the resulting sensitivity.}
\renewcommand{\arraystretch}{1.23} 
  \begin{tabular}{c||c|c|c|c} 
     & Case1 & Case2 & Case3 & Case4 \\ \hline
    $N$ & \multicolumn{2}{|c|}{$2$} & \multicolumn{2}{|c}{$5$} \\  \hline
    $M_1$ [dB]  & 45 & 30 & 45  & 30 \\  \hline
    $x$ & 0.90 & 0.78 & 0.53  & 0.393 \\  \hline
    $\beta_{\scriptscriptstyle A}$ & 0.2 & 0.1 & 0.07 & 0.03 \\ \hline
    $\beta_{\scriptscriptstyle B}$ & $-0.0412$ & $-0.0291$ & 0.0034 & 0.0046 \\ \hline
    $S_{A}$ & 0.3388 & 0.7259 & 1.0718 & 1.4094 \\ \hline
    $S_{B}$ & 0.1190 & 0.5271 & 0.7428 & 1.2802 \\ \hline
    $g_{m}$ [dB] & 8.1310 & 18.4593 & 8.5699 & 19.9847 \\ \hline
  \end{tabular}
\label{table_parameterset-BS}
\end{center}
\end{table}

The superiority of the type-B scheme over the type-A is guaranteed to hold only at the 
center frequency $\omega=0$. 
Thus, in this section, we focus on a specific system and numerically investigate the 
frequency dependence of amplification gain in those two schemes, with particular 
attention to the robustness and stability properties.

The amplifier is the ideal NDPA discussed in Section~\ref{sec:2.2}. 
The controller is a partially transmissive mirror called the {\it beam-splitter (BS)}, which is 
a 2-inputs and 2-outputs passive static system with the following transfer function matrix: 
\[
    K_\bullet(s) =\left[\begin{array}{cc}
                \alpha_\bullet & -\beta_\bullet \\  
                \beta_\bullet & \alpha_\bullet  \\
              \end{array}\right ], ~~\alpha_\bullet^2+\beta_\bullet^2=1,
\]
where $\bullet=A, B$. 
The real parameters $\alpha_\bullet, \beta_\bullet\in{\mathbb R}$ represent the 
transmissivity and reflectivity of the mirror, respectively. 
Note that, from Eq.~\eqref{closed-loop G general}, the single NDPA with this controller 
has the amplification gain $1/|\beta_\bullet|$ in the limit $x \to 1-0$.

We consider the four cases summarized in Table~\ref{table_parameterset-BS}; 
the number of amplifiers is $N=2$ (Cases 1 and 2) or $N=5$ (Cases 3 and 4); 
the gain of the (1,1) element of $[G(0)]^N$, i.e., the non-controlled cascaded NDPA at 
$\omega=0$, is $M_1=45$ dB (Cases 1 and 3) or $M_1=30$ dB (Cases 2 and 4). 
In each case the cavity decay rate of the NDPA is fixed to $\kappa=1.8 \times 10^7$ Hz 
\cite{Iida2012,Mabuchi2013}, while $x=2\varepsilon/\kappa$ is chosen so that $M_1$ 
equals to 45 dB or 30 dB. 
The reflectivity $\beta_B$ was determined as follows; 
first we fix the parameters of the type-A system, $x$ and $\beta_A$, and then 
$\beta_B$ is determined so that the gains at $\omega=0$ of both of the schemes 
are the same, i.e., $|G^{\rm fb}_{1A}|=|G^{\rm fb}_{1B}|$.

First, let us see the stability of the feedback-controlled system. 
For the type-A system, it is enough to analyze the stability of the single feedback-controlled NDPA; 
its characteristic equation is given by 
\begin{align*}
           s^2 +\frac{\kappa}{1-\beta_A} s 
                     +\frac{1+\beta_A}{1-\beta_A}\frac{\,\,\kappa^2}{4}-\varepsilon^2=0.
\end{align*}
The system is stable if and only if both of the two solutions $s$ of this equation satisfy 
${\rm Re}(s)<0$, which leads to 
$x=2\varepsilon/\kappa < \sqrt{(1+\beta_A)/(1-\beta_A)}$. 
This condition is always satisfied if the NDPA is stable ($x <1$) and 
$0 \leq \beta_A< 1$.

\begin{figure}[!t]
\begin{center}
\includegraphics[width=0.95\hsize]{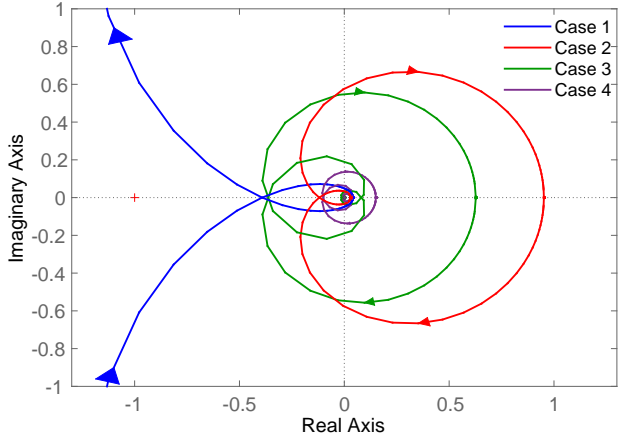}
\caption{Nyquist plots of the type-B controlled system.}
\label{fig_Nyquist-BS}
\end{center}
\end{figure}

To analyze the stability property of the type-B system, we use the {\it Nyquist plot}, which is now 
directly applicable because all the parameters ($\kappa, \varepsilon, \alpha_B, \beta_B$) are real. 
The Nyquist plot is the vector plot of the open-loop transfer function $L(s)$, i.e., 
the trajectory of $({\rm Re}\{L(i \omega)\}, {\rm Im}\{L(i \omega)\})$ with $\omega\in(-\infty, +\infty)$; 
note that $L(s)=G(s)K(s)$ for the classical system \eqref{eq_classicalFBamp}. 
The feedback-controlled system is stable if and only if there is no encirclement of the point 
$(-1,0)$, provided that $L(s)$ has no unstable poles. 
Now, from Eq.~\eqref{closed-loop G general} the type-B system has the open-loop transfer 
function $L(s) = - [G^N]_{22}(s) \beta_B$, where $[G^N]_{22}(s)$ is the (2,2) element of $G(s)^N$. 
The Nyquist plots are shown in Fig.~\ref{fig_Nyquist-BS}; 
hence, from the above stability criterion, the type-B system is stable in all Cases.

\begin{figure}[!t]
    \begin{tabular}{cc}
      \begin{minipage}{0.47\hsize}
        \centering
        \includegraphics[width=1.0\hsize]{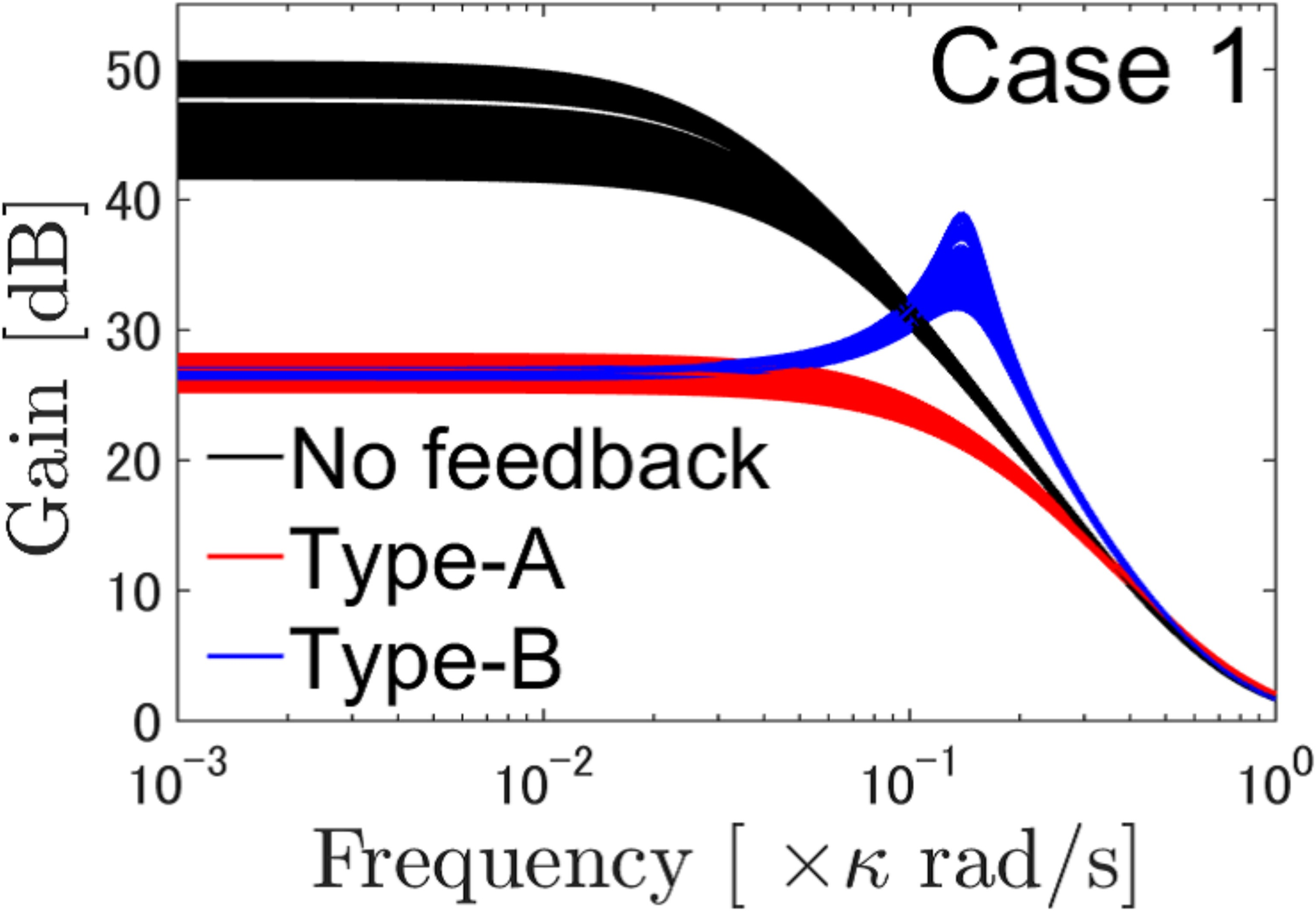}
        \subcaption{$N=2,\,M_1 \approx 45\,{\rm [dB]}$}
        \label{fig_Gainplot-BS-case1}
      \end{minipage} 
      \begin{minipage}{0.47\hsize}
        \centering
        \includegraphics[width=1.0\hsize]{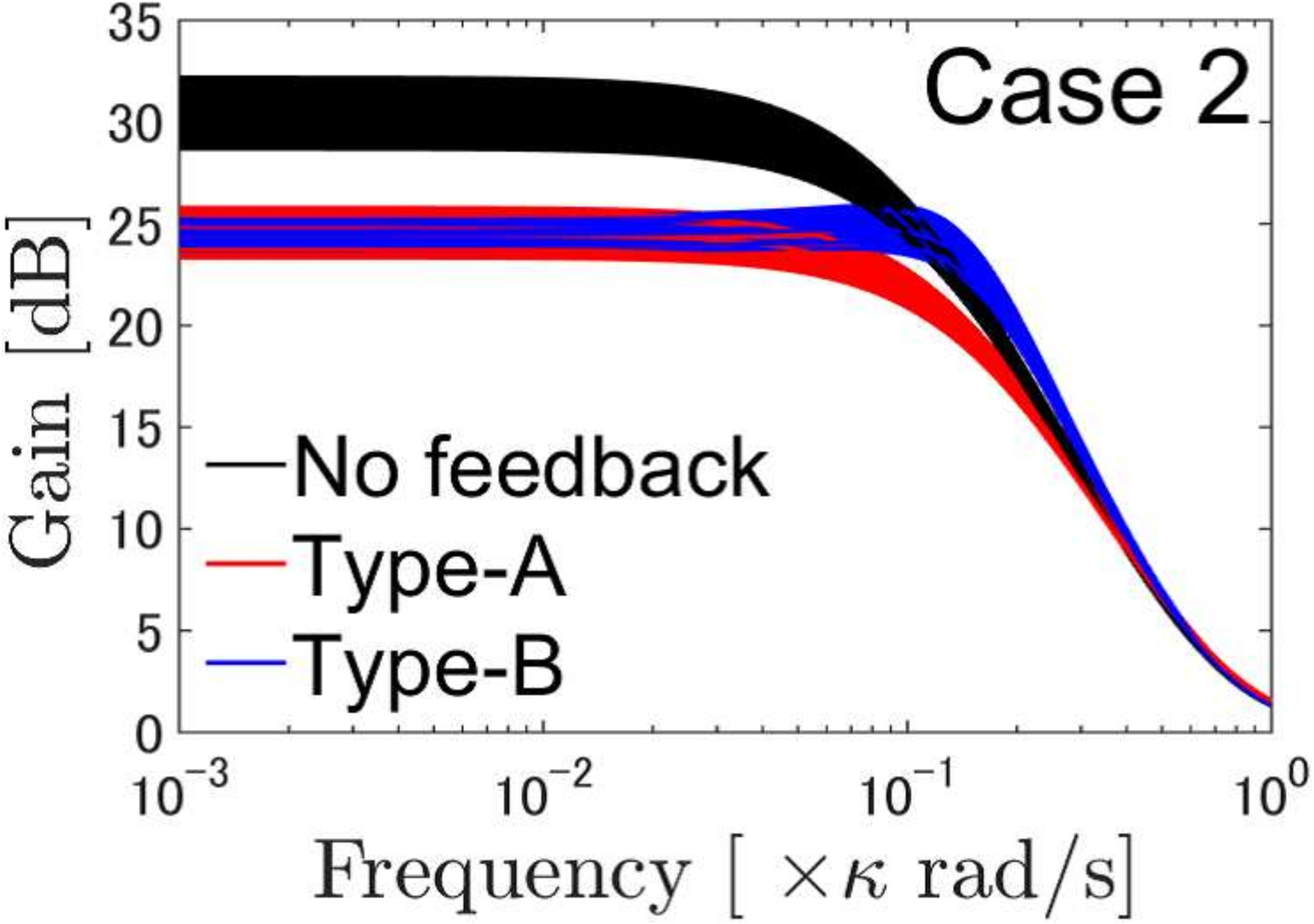}
        \subcaption{$N=2,\,M_1 \approx 30\,{\rm [dB]}$}
        \label{fig_Gainplot-BS-case2}
      \end{minipage} \\
      \begin{minipage}{1\hsize}
        \vspace{5mm}
      \end{minipage} \\
      \begin{minipage}{0.47\hsize}
        \centering
        \includegraphics[width=1.0\hsize]{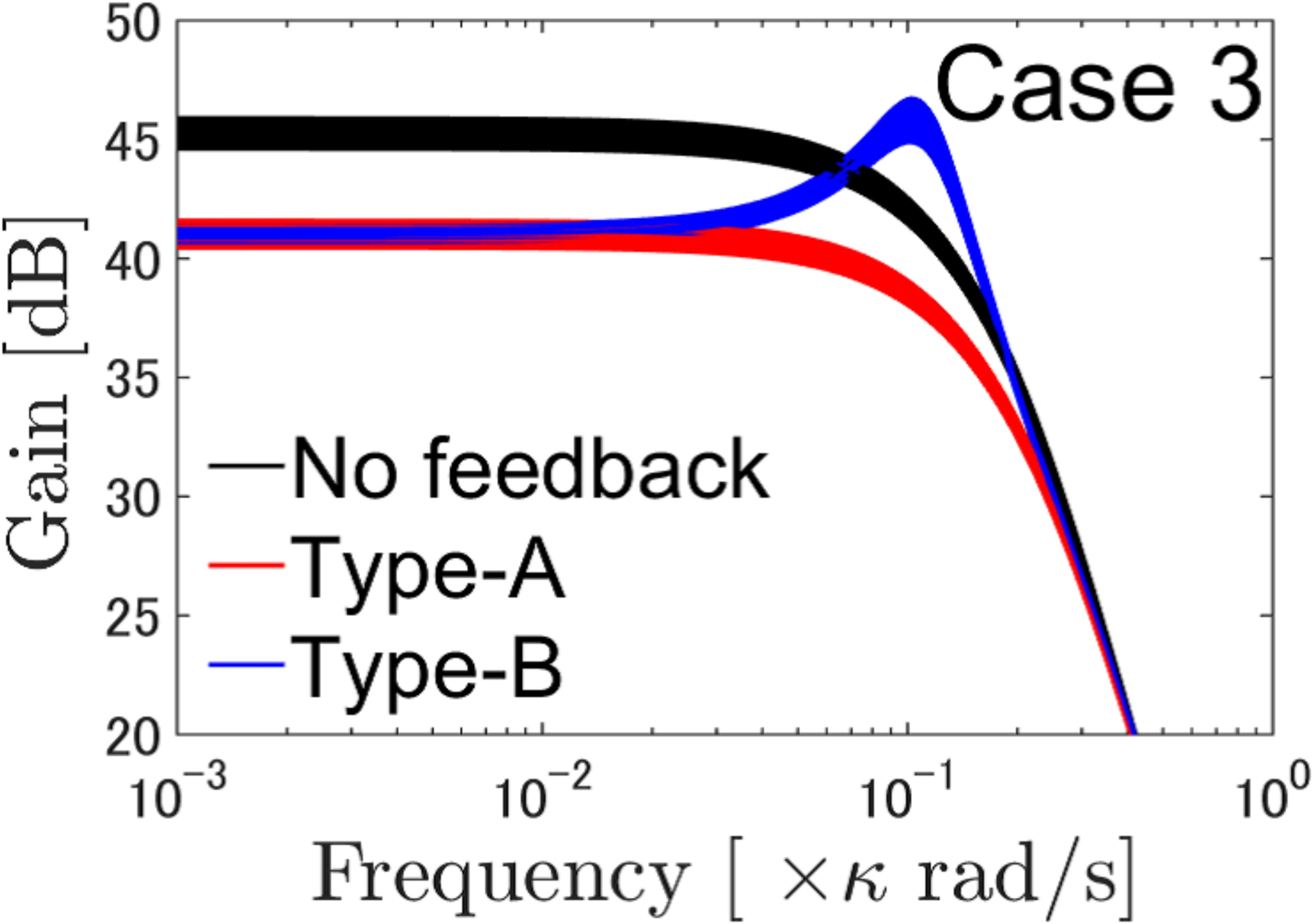}
        \subcaption{$N=5,\,M_1 \approx 45\,{\rm [dB]}$}
        \label{fig_Gainplot-BS-case3}
      \end{minipage} 
      \begin{minipage}{0.47\hsize}
        \centering
        \includegraphics[width=1.0\hsize]{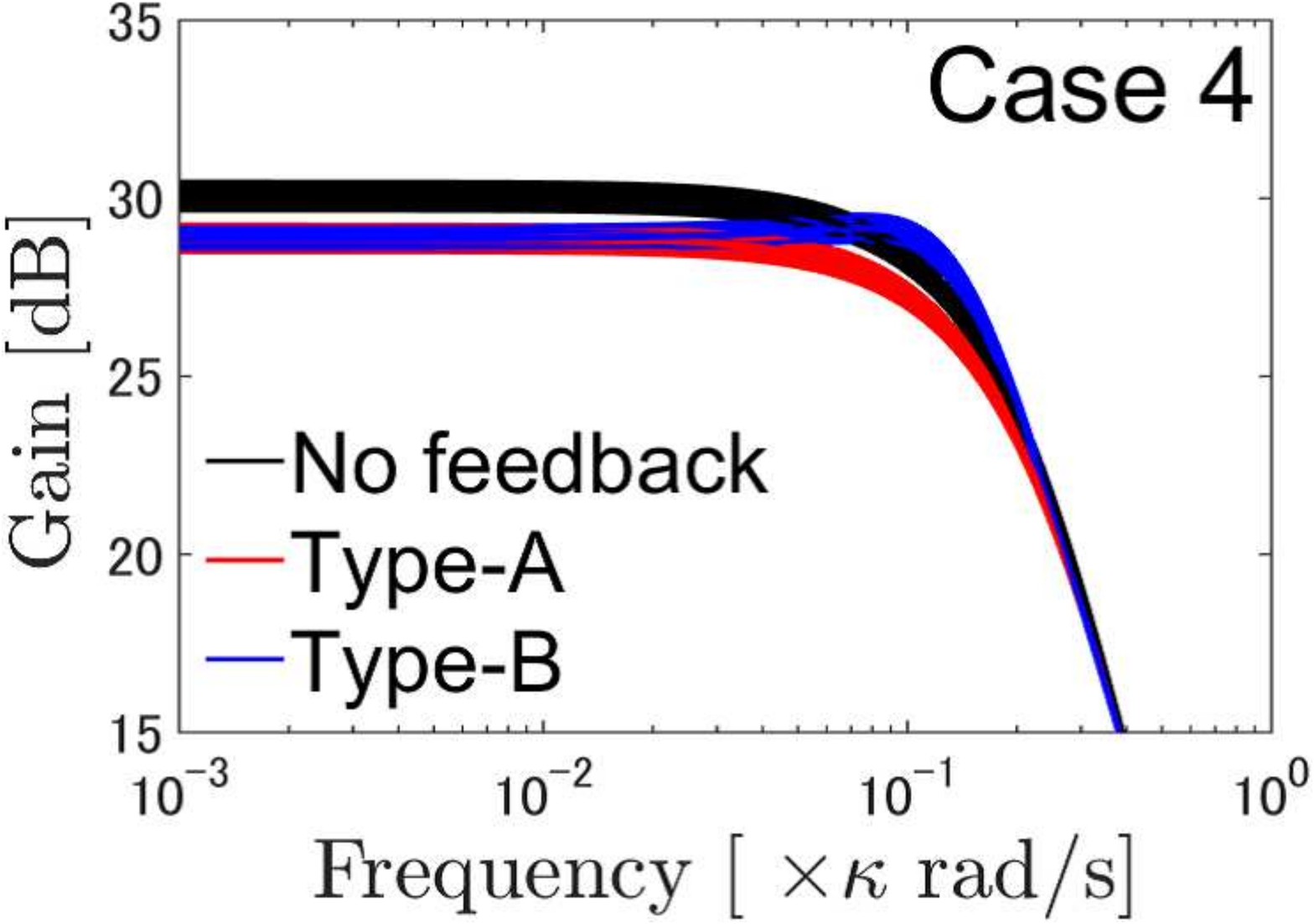}
        \subcaption{$N=5,\,M_1 \approx 30\,{\rm [dB]}$}
        \label{fig_Gainplot-BS-case4}
      \end{minipage}
   \end{tabular}
\caption{Gain plots of the feedback-controlled system.}
\label{fig_GainplotBS}
\end{figure}

Next we discuss the sensitivity. 
To see this explicitly, suppose that the characteristic change of the amplifier, $\Delta G_1$, 
stems from the fluctuation of the parameter $\varepsilon$. 
We model this uncertainty as $\varepsilon'=(1+0.05r)\varepsilon$, where $r$ is the random 
number generated from the uniform distribution over $[-1, 1]$; 
that is, the nominal parameter $x=2\varepsilon/\kappa$ given in Table~\ref{table_parameterset-BS} 
experiences up to 5$\%$ deviation. 
The gain plots are shown in Fig.~\ref{fig_GainplotBS}, where the red and blue lines represent 
the gains of the type-A and the type-B systems, respectively. 
Also the black lines are the gain plots of the cascaded amplifier without feedback. 
In each scheme (color), 100 sample paths are produced from the above-mentioned 
probability distribution. 
The figure shows that, in all Cases, the gain fluctuation of the controlled systems at $\omega=0$ 
are smaller than that of the uncontrolled system; 
that is, the feedback control always works well to suppress the gain fluctuation of the amplifier, 
at the price of decreasing the gain. 
Moreover, the fluctuation of the gain at $\omega=0$ of the type-B controlled system is 
always smaller than that of the type-A, i.e., $|S_B| < |S_A|$, as proven in Section~\ref{sec:3}. 
However, importantly, this fact does not hold over all frequencies; 
in particular in Cases 1 and 3, the type-A scheme is better than the type-B, at the frequency 
$\omega\sim \kappa/10$ where there is a peaking in the gain.

Finally we discuss the control performance, with the focus on both stability and sensitivity. 
The Nyquist plot can be used to quantify how much the system is stable, in terms of the 
gain margin $g_{m}=1/|L(i \omega_{pc})|$ with $\omega_{pc}$ the 
phase crossover frequency satisfying $\angle L(i \omega_{pc}) =-180^\circ$. 
Now, as shown in Table~\ref{table_parameterset-BS}, the gain margin $g_m$ in Cases 1 
and 3 are smaller than that in Cases 2 and 4. 
Hence the systems in Cases 1 and 3 are less stable than those in Cases 2 and 4; 
actually a peaking appears in Figs.~\ref{fig_GainplotBS}(a) and (c), but not in (b) and (d). 
However, as implied by Fig.~\ref{fig_GainplotBS}, it is harder to reduce the sensitivity in 
Cases 2 and 4, compared to Cases 1 and 3. 
That is, there is a tradeoff between the stability and robustness. 
Note also that the controlled system with less number of amplifiers has the better sensitivity; 
in fact the controlled system composed of $N=5$ amplifiers, which yet has the same level 
of sensitivity as that of the system with $N=2$, is often unstable.


\section{Concluding remark}
\label{sec:5}

The long-term goal of this work is to develop the design theory for feedback-controlled 
quantum networks containing amplifiers, corresponding to the established classical one 
\cite{Graeme1971}--\cite{Horrocks1990}. 
Toward this goal, as an important first step, this letter gives the following theorem: 
to construct a robust high-gain quantum amplifier from some low-gain amplifiers, it is always 
better to stabilize the cascaded amplifier via a single feedback controller, than to take a 
cascade connection of feedback-controlled amplifiers. 
Recall that, although this is the same conclusion as the classical one, the proof of this fact 
is highly non-trivial. 
Also, as stated in Remark~1 and shown in Section~\ref{sec:4}, the sensitivity functions of the 
quantum feedback amplifiers have different characteristic from the classical counterparts 
in robustness and stability. 
As a consequence, a more careful sensitivity analysis will be required in general for designing 
a practical quantum network device, e.g., a robust quantum communication channel over 
a specific bandwidth 
\cite{Pirandola2017}--\cite{Elemy2017}.



\appendix


First we prove Eq.~\eqref{eq_fluctuation relationship}. 
If the gain of the $j$-th amplifier changes as $G_{1} \to G_{1} +\Delta G_{1}$ and 
$G_{2} \to G_{2} +\Delta G_{2}$, then 
$M_{1} = (\lambda_{+}^N + \lambda_{-}^N)/2$ 
changes as follows;
\begin{align*}
\Delta M_{1}
&=\frac{1}{2}\,\Bigl[
(G_{1}+\Delta G_{1}+G_{2}+\Delta G_{2})\lambda_{+}^{N-1} \Bigr. \\
&~~~~~~~~~\Bigl. +(G_{1}+\Delta G_{1}-G_{2}-\Delta G_{2})\lambda_{-}^{N-1}\Bigr] -M_{1} \\
&=\frac{1}{2}\,\Bigl[
(\Delta G_{1}+\Delta G_{2})\lambda_{+}^{N-1} + (\Delta G_{1}-\Delta G_{2})\lambda_{-}^{N-1} \Bigr] \\
&=\frac{\Delta G_{1}}{2}\Biggl[
\left(1+\frac{G_{1}}{G_{2}}\right)\lambda_{+}^{N-1} +\left(1-\frac{G_{1}}{G_{2}}\right)\lambda_{-}^{N-1} \Biggr] \\ 
&=\frac{\Delta G_{1}}{2G_{2}}\Bigl[ \lambda_{+}^N - \lambda_{-}^N \Bigr] =\frac{M_{2}}{G_{2}}\,\Delta G_{1}.
\end{align*}

Next we prove Eq.~\eqref{main result}. 
To make a fair comparison, we assume that both the controlled systems have the same 
amplification gain at $\omega=0$, i.e., $|G^{\rm fb}_{1A}|=|G^{\rm fb}_{1B}|$, 
which leads to $|G^{\rm fb}_{2A}|=|G^{\rm fb}_{2B}|$. 
Then we have 
\[
     \frac{|S_{B}|}{|S_{A}|} = 
            \left| \frac{K_{1B}}{K_{1A}} \frac{1+G_{1}K_{2A}}{1+M_{1}K_{2B}} \right| 
    =\left|\frac{1+G_{1}K_{2A}}{K_{1A}}\right| 
      \left|\frac{\left(\lambda^{\rm fb}_{+}\right)^N 
                            - \left(\lambda^{\rm fb}_{-}\right)^N}
                    {\lambda_{+}^N - \lambda_{-}^N} \right|.
\]
Here, from the relations $a^{n}-b^{n} = (a-b)\sum_{k=1}^n a^{n-k} b^{k-1}$ and 
$\lambda_{+} \lambda_{-} =
\lambda^{\rm fb}_{+} \lambda^{\rm fb}_{-}=1$, we have
\begin{align*}
&\hspace{-0.4cm}
\left(\lambda^{\rm fb}_{+}\right)^N - \left(\lambda^{\rm fb}_{-}\right)^N \\
=&\left(\lambda^{\rm fb}_{+}-\lambda^{\rm fb}_{-}\right)
\Bigl[\left(\lambda^{\rm fb}_{+}\right)^{N-1}+\left(\lambda^{\rm fb}_{+}\right)^{N-3}+ \cdots +\left(\lambda^{\rm fb}_{+}\right)^{-(N-1)} \Bigr] \\
=& \frac{2G_{2}K_{1A}}{1+G_{1}K_{2A}} {\sum^{N}_{k=1} \left(\lambda^{\rm fb}_{+}\right)^{N-2k+1}},
\end{align*}
and likewise 
$\lambda_{+}^N - \lambda_{-}^N 
= 2G_{2}{\sum^{N}_{k=1} \lambda_{+}^{N-2k+1}}$. 
Hence, $|S_{B}|/|S_{A}|$ is now expressed as 
\begin{align}
    \frac{|S_{B}|}{|S_{A}|}
        =\left| \frac{\sum^{N}_{k=1} \left( \lambda^{\rm fb}_{+} \right)^{N-2k+1}}
                         {\sum^{N}_{k=1} \lambda_{+}^{N-2k+1}} \right|.
\label{eq_Qindex2}
\end{align}
In addition to the condition $|G^{\rm fb}_{1A}|=|G^{\rm fb}_{1B}|$, we assume that 
the gains of both of the type-A and type-B controlled systems are smaller than the gain of 
the non-controlled cascaded amplifier; 
$|G^{\rm fb}_{1A}|=|G^{\rm fb}_{1B}| < |M_1|$, which is represented as 
\[
    \frac{|G^{\rm fb}_{1A}|}{|M_1|}
          =\left| \frac{\left( \lambda^{\rm fb}_{+} \right)^{k}
                                                +\left( \lambda^{\rm fb}_{+} \right)^{-k}}
                           { \lambda_{+}^{k}
                                                + \lambda_{+}^{-k}}\right| < 1, ~~
     \forall k=1, \cdots, N.
\]
Then, if $N$ is odd, Eq.~\eqref{eq_Qindex2} leads to
\begin{align*}
    \frac{|S_{B}|}{|S_{A}|}
        &= \left| \frac{1+ \sum^{(N-1)/2}_{k=1} \left[
                    \left( \lambda^{\rm fb}_{+} \right)^{2k}
        + \left( \lambda^{\rm fb}_{+} \right)^{-2k}  \right]  }
                            {1+ \sum^{(N-1)/2}_{k=1} \left(
                                 \lambda_{+}^{2k}
                                     + \lambda_{+}^{-2k}   \right) }  \right| \\
        &= \frac{1+ \sum^{(N-1)/2}_{k=1} \left[
                    \left( \lambda^{\rm fb}_{+} \right)^{2k}
        + \left( \lambda^{\rm fb}_{+} \right)^{-2k}  \right] }
                            {1+ \sum^{(N-1)/2}_{k=1} \left(
                                 \lambda_{+}^{2k}
                                      + \lambda_{+}^{-2k}  \right) } <1. 
\end{align*}
Also, if $N$ is even, particularly $N=4l-2 ~ (l=1,2, \cdots)$, 
\begin{align*}
    \hspace{-0.1cm}
    \frac{|S_{B}|}{|S_{A}|} 
    &= \left| 
       \frac{\lambda^{\rm fb}_{+} +\left( \lambda^{\rm fb}_{+} \right)^{-1}}
              {\lambda_{+} + \lambda_{+}^{-1}} \right|  
       \frac{1 + \sum^{l-1}_{k=1} \left[ 
                    \left( \lambda^{\rm fb}_{+} \right)^{4k}
                    + \left( \lambda^{\rm fb}_{+} \right)^{-4k}  \right]  }
                            {1+ \sum^{l-1}_{k=1} \left( 
                                 \lambda_{+}^{4k}
                                 + \lambda_{+}^{-4k} \right) }, 
\end{align*}
and if $N=4l ~ (l=1,2, \cdots)$, 
\begin{align*}
    \hspace{-0.1cm}
    \frac{|S_{B}|}{|S_{A}|} 
    &= \left| 
       \frac{\lambda^{\rm fb}_{+} +\left( \lambda^{\rm fb}_{+} \right)^{-1}}
              {\lambda_{+} + \lambda_{+}^{-1}} \right|  
      \frac{ \sum^{l}_{k=1} \left[ 
                    \left( \lambda^{\rm fb}_{+} \right)^{4k-2}
                    + \left( \lambda^{\rm fb}_{+} \right)^{-(4k-2)}  \right]  }
                            { \sum^{l}_{k=1} 
                                 \left[\lambda_{+}^{4k-2}
                                 + \lambda_{+}^{-(4k-2)} \right] 
}, 
\end{align*}
which are both less than $1$. 
This completes the proof.



\end{document}